\documentclass[aps,prl,twocolumn,superscriptaddress,showpacs]{revtex4} 

\usepackage{graphicx}
\usepackage{dcolumn}
\usepackage{bm}
\usepackage{lineno}
\usepackage{setspace}
\usepackage{amsfonts}
\usepackage{amssymb,amsmath}
\usepackage{xspace}
\usepackage{color}

\newcommand{\ppbar}  {\ensuremath{p\bar{p}}\xspace}

\newcommand{\ttbar}  {\ensuremath{t\bar{t}}\xspace}

\newcommand{\met}{\mbox{${\not\!\!E_T}$}}

\def\ttbar {\ensuremath{t\overline t}}
\def\ppbar {\ensuremath{p\overline p}}
\def\Cts   {\ensuremath{\cos \theta^{*}}}
\def\CtsSq   {\ensuremath{\cos^{2} \theta^{*}}}

\begin{document}


\title{Measurement of $W$-Boson Polarization in Top-quark Decay using the Full CDF Run~II Data Set}

\affiliation{Institute of Physics, Academia Sinica, Taipei, Taiwan 11529, Republic of China}
\affiliation{Argonne National Laboratory, Argonne, Illinois 60439, USA}
\affiliation{University of Athens, 157 71 Athens, Greece}
\affiliation{Institut de Fisica d'Altes Energies, ICREA, Universitat Autonoma de Barcelona, E-08193, Bellaterra (Barcelona), Spain}
\affiliation{Baylor University, Waco, Texas 76798, USA}
\affiliation{Istituto Nazionale di Fisica Nucleare Bologna, $^{ee}$University of Bologna, I-40127 Bologna, Italy}
\affiliation{University of California, Davis, Davis, California 95616, USA}
\affiliation{University of California, Los Angeles, Los Angeles, California 90024, USA}
\affiliation{Instituto de Fisica de Cantabria, CSIC-University of Cantabria, 39005 Santander, Spain}
\affiliation{Carnegie Mellon University, Pittsburgh, Pennsylvania 15213, USA}
\affiliation{Enrico Fermi Institute, University of Chicago, Chicago, Illinois 60637, USA}
\affiliation{Comenius University, 842 48 Bratislava, Slovakia; Institute of Experimental Physics, 040 01 Kosice, Slovakia}
\affiliation{Joint Institute for Nuclear Research, RU-141980 Dubna, Russia}
\affiliation{Duke University, Durham, North Carolina 27708, USA}
\affiliation{Fermi National Accelerator Laboratory, Batavia, Illinois 60510, USA}
\affiliation{University of Florida, Gainesville, Florida 32611, USA}
\affiliation{Laboratori Nazionali di Frascati, Istituto Nazionale di Fisica Nucleare, I-00044 Frascati, Italy}
\affiliation{University of Geneva, CH-1211 Geneva 4, Switzerland}
\affiliation{Glasgow University, Glasgow G12 8QQ, United Kingdom}
\affiliation{Harvard University, Cambridge, Massachusetts 02138, USA}
\affiliation{Division of High Energy Physics, Department of Physics, University of Helsinki and Helsinki Institute of Physics, FIN-00014, Helsinki, Finland}
\affiliation{University of Illinois, Urbana, Illinois 61801, USA}
\affiliation{The Johns Hopkins University, Baltimore, Maryland 21218, USA}
\affiliation{Institut f\"{u}r Experimentelle Kernphysik, Karlsruhe Institute of Technology, D-76131 Karlsruhe, Germany}
\affiliation{Center for High Energy Physics: Kyungpook National University, Daegu 702-701, Korea; Seoul National University, Seoul 151-742, Korea; Sungkyunkwan University, Suwon 440-746, Korea; Korea Institute of Science and Technology Information, Daejeon 305-806, Korea; Chonnam National University, Gwangju 500-757, Korea; Chonbuk National University, Jeonju 561-756, Korea; Ewha Womans University, Seoul, 120-750, Korea}
\affiliation{Ernest Orlando Lawrence Berkeley National Laboratory, Berkeley, California 94720, USA}
\affiliation{University of Liverpool, Liverpool L69 7ZE, United Kingdom}
\affiliation{University College London, London WC1E 6BT, United Kingdom}
\affiliation{Centro de Investigaciones Energeticas Medioambientales y Tecnologicas, E-28040 Madrid, Spain}
\affiliation{Massachusetts Institute of Technology, Cambridge, Massachusetts 02139, USA}
\affiliation{Institute of Particle Physics: McGill University, Montr\'{e}al, Qu\'{e}bec H3A~2T8, Canada; Simon Fraser University, Burnaby, British Columbia V5A~1S6, Canada; University of Toronto, Toronto, Ontario M5S~1A7, Canada; and TRIUMF, Vancouver, British Columbia V6T~2A3, Canada}
\affiliation{University of Michigan, Ann Arbor, Michigan 48109, USA}
\affiliation{Michigan State University, East Lansing, Michigan 48824, USA}
\affiliation{Institution for Theoretical and Experimental Physics, ITEP, Moscow 117259, Russia}
\affiliation{University of New Mexico, Albuquerque, New Mexico 87131, USA}
\affiliation{The Ohio State University, Columbus, Ohio 43210, USA}
\affiliation{Okayama University, Okayama 700-8530, Japan}
\affiliation{Osaka City University, Osaka 588, Japan}
\affiliation{University of Oxford, Oxford OX1 3RH, United Kingdom}
\affiliation{Istituto Nazionale di Fisica Nucleare, Sezione di Padova-Trento, $^{ff}$University of Padova, I-35131 Padova, Italy}
\affiliation{University of Pennsylvania, Philadelphia, Pennsylvania 19104, USA}
\affiliation{Istituto Nazionale di Fisica Nucleare Pisa, $^{gg}$University of Pisa, $^{hh}$University of Siena and $^{ii}$Scuola Normale Superiore, I-56127 Pisa, Italy, $^{mm}$INFN Pavia and University of Pavia, I-27100 Pavia, Italy}
\affiliation{University of Pittsburgh, Pittsburgh, Pennsylvania 15260, USA}
\affiliation{Purdue University, West Lafayette, Indiana 47907, USA}
\affiliation{University of Rochester, Rochester, New York 14627, USA}
\affiliation{The Rockefeller University, New York, New York 10065, USA}
\affiliation{Istituto Nazionale di Fisica Nucleare, Sezione di Roma 1, $^{jj}$Sapienza Universit\`{a} di Roma, I-00185 Roma, Italy}
\affiliation{Texas A\&M University, College Station, Texas 77843, USA}
\affiliation{Istituto Nazionale di Fisica Nucleare Trieste/Udine; $^{nn}$University of Trieste, I-34127 Trieste, Italy; $^{kk}$University of Udine, I-33100 Udine, Italy}
\affiliation{University of Tsukuba, Tsukuba, Ibaraki 305, Japan}
\affiliation{Tufts University, Medford, Massachusetts 02155, USA}
\affiliation{University of Virginia, Charlottesville, Virginia 22906, USA}
\affiliation{Waseda University, Tokyo 169, Japan}
\affiliation{Wayne State University, Detroit, Michigan 48201, USA}
\affiliation{University of Wisconsin, Madison, Wisconsin 53706, USA}
\affiliation{Yale University, New Haven, Connecticut 06520, USA}

\author{T.~Aaltonen}
\affiliation{Division of High Energy Physics, Department of Physics, University of Helsinki and Helsinki Institute of Physics, FIN-00014, Helsinki, Finland}
\author{S.~Amerio}
\affiliation{Istituto Nazionale di Fisica Nucleare, Sezione di Padova-Trento, $^{ff}$University of Padova, I-35131 Padova, Italy}
\author{D.~Amidei}
\affiliation{University of Michigan, Ann Arbor, Michigan 48109, USA}
\author{A.~Anastassov$^x$}
\affiliation{Fermi National Accelerator Laboratory, Batavia, Illinois 60510, USA}
\author{A.~Annovi}
\affiliation{Laboratori Nazionali di Frascati, Istituto Nazionale di Fisica Nucleare, I-00044 Frascati, Italy}
\author{J.~Antos}
\affiliation{Comenius University, 842 48 Bratislava, Slovakia; Institute of Experimental Physics, 040 01 Kosice, Slovakia}
\author{G.~Apollinari}
\affiliation{Fermi National Accelerator Laboratory, Batavia, Illinois 60510, USA}
\author{J.A.~Appel}
\affiliation{Fermi National Accelerator Laboratory, Batavia, Illinois 60510, USA}
\author{T.~Arisawa}
\affiliation{Waseda University, Tokyo 169, Japan}
\author{A.~Artikov}
\affiliation{Joint Institute for Nuclear Research, RU-141980 Dubna, Russia}
% ...

\collaboration{CDF Collaboration\footnote{With visitors from
$^a$Istituto Nazionale di Fisica Nucleare, Sezione di Cagliari, 09042 Monserrato (Cagliari), Italy,
$^b$University of California Irvine, Irvine, CA 92697, USA,
$^c$University of California Santa Barbara, Santa Barbara, CA 93106, USA,
$^d$University of California Santa Cruz, Santa Cruz, CA 95064, USA,
$^e$Institute of Physics, Academy of Sciences of the Czech Republic, 182~21, Czech Republic,
$^f$CERN, CH-1211 Geneva, Switzerland,
$^g$Cornell University, Ithaca, NY 14853, USA,
$^h$University of Cyprus, Nicosia CY-1678, Cyprus,
$^i$Office of Science, U.S. Department of Energy, Washington, DC 20585, USA,
$^j$University College Dublin, Dublin 4, Ireland,
$^k$ETH, 8092 Z\"{u}rich, Switzerland,
$^l$University of Fukui, Fukui City, Fukui Prefecture, Japan 910-0017,
$^m$Universidad Iberoamericana, Lomas de Santa Fe, M\'{e}xico, C.P. 01219, Distrito Federal,
$^n$University of Iowa, Iowa City, IA 52242, USA,
$^o$Kinki University, Higashi-Osaka City, Japan 577-8502,
$^p$Kansas State University, Manhattan, KS 66506, USA,
$^q$Brookhaven National Laboratory, Upton, NY 11973, USA,
$^r$University of Manchester, Manchester M13 9PL, United Kingdom,
$^s$Queen Mary, University of London, London, E1 4NS, United Kingdom,
$^t$University of Melbourne, Victoria 3010, Australia,
$^u$Muons, Inc., Batavia, IL 60510, USA,
$^v$Nagasaki Institute of Applied Science, Nagasaki 851-0193, Japan,
$^w$National Research Nuclear University, Moscow 115409, Russia,
$^x$Northwestern University, Evanston, IL 60208, USA,
$^y$University of Notre Dame, Notre Dame, IN 46556, USA,
$^z$Universidad de Oviedo, E-33007 Oviedo, Spain,
$^{aa}$CNRS-IN2P3, Paris, F-75205 France,
$^{bb}$Texas Tech University, Lubbock, TX 79609, USA,
$^{cc}$Universidad Tecnica Federico Santa Maria, 110v Valparaiso, Chile,
$^{dd}$Yarmouk University, Irbid 211-63, Jordan,
$^{ll}$Universite catholique de Louvain, 1348 Louvain-La-Neuve, Belgium,
$^{oo}$University of Z\"{u}rich, 8006 Z\"{u}rich, Switzerland,
$^{pp}$Massachusetts General Hospital and Harvard Medical School, Boston, MA 02114 USA,
$^{qq}$Hampton University, Hampton, VA 23668, USA,
$^{rr}$Los Alamos National Laboratory, Los Alamos, NM 87544, USA
}}
\noaffiliation

\date{\today}

\begin{abstract}
We measure the polarization of $W$ bosons from top-quark ($t$) decays into
final states with a charged lepton and jets, $\ttbar \to W^+ b W^- \bar{b} \to \ell \nu b q \bar{q}' \bar{b}$, using the full Run~II data set collected by the CDF~II detector, corresponding to an integrated luminosity of 8.7~$\rm{fb}^{-1}$. A model-independent method simultaneously determines the fraction of longitudinal ($f_0$) and right-handed ($f_+$) $W$ bosons to yield
$f_{0} = 0.726  \pm 0.066\: (\rm{stat}) \pm 0.067\: (\rm{syst})$ and
$f_{+} = -0.045 \pm 0.044\: (\rm{stat}) \pm 0.058\: (\rm{syst})$
with a correlation coefficient of --0.69.
Additional results are presented under various standard model 
assumptions.
No significant discrepancies with the standard model are observed.
\end{abstract}



\maketitle


The top quark was first observed in the Tevatron Run~I (1992-1996) data sets collected by the CDF and D0 experiments at Fermilab~\cite{discovery}. Because the mass of the top quark is large, beyond-the-standard-model (BSM) physics contributions can affect the top-quark phenomenology in a wide variety of ways: the production mechanisms can be affected, the decay widths can be altered, its intrinsic properties can be changed, and the experimental signature can be mimicked by a new particle of similar mass.  Thus a principal goal of the Tevatron Run~II (2001-2011) program, which produced data samples {\cal O}(100) times larger than Run~I, was to thoroughly explore the properties of the top quark.  This Letter reports a measurement of the $W$-boson polarization from top-quark decay using the full Run~II data set collected by the CDF~II experiment.  The focus is on 
the lepton-plus-jets final state, 
$\ttbar \to W^+ b W^- \bar{b} \to \ell \nu b q \bar{q}' \bar{b}$, which 
provides the most sensitive determination of the $W$ polarization due to 
its high yield, low background, and constrained kinematics.  
At present the most precise experimental knowledge of the $W$-boson polarization comes from the ATLAS experiment~\cite{AtlasResult} and from the combination of existing Tevatron results~\cite{TevWhelComb}. The results reported here supersede the previous CDF measurements in Ref.~\cite{CDF2.7ifb} and have a total uncertainty comparable to the Tevatron combination and a statistical precision a factor of 1.6 smaller than Ref.~\cite{CDF2.7ifb}.

The top quark~\cite{ChargeConjugate} almost always decay to a $W^+$ boson and a $b$ quark~\cite{PDG2010} via the charged-current weak interaction whose $V-A$ structure in the standard model (SM) specifies the $tWb$ coupling and the resulting $W$-boson polarization. Due to its large mass, the top quark decays before hadronizing and thus offers a direct probe of the $tWb$ coupling.  At first order in the SM perturbative expansion~\cite{WtbCoupling}, the $W^+$ boson is expected to have longitudinal polarization $f_{0} = 0.696$, left-handed polarization $f_- = 0.303$, and right-handed polarization $f_+ = 3.8 \times 10^{-4}$ for a top-quark mass $m_t = 172.5$ GeV/$c^2$~\cite{mtop}, a $b$-quark mass $m_b = 4.79$  GeV/$c^2$~\cite{PDG2010}, and a $W$-boson mass $M_W = 80.413$ GeV/$c^2$~\cite{WmassCDFrunII}. Higher-order quantum chromodynamic and electroweak radiative corrections, as well as the uncertainties on the values of $m_t$, $m_b$, and $M_W$, change these predictions at the 1-2\% relative level~\cite{PDG2010, QCDandEWcorr}.  The presence of anomalous couplings at the $tWb$ vertex, due to contributions from BSM physics, can modify the observed $W$ polarization with respect to the SM expectations~\cite{WtbCoupling}.


In this Letter, three different measurements of the $W$-boson polarization are performed: 
a model-independent determination that simultaneously
measures $f_0$ and $f_+$, 
a measurement of $f_0$ for fixed $f_+$=0, which enhances sensitivity to
anomalous tensor couplings, and 
a measurement of $f_+$ for fixed $f_0=0.70$, which enhances sensitivity to anomalous right-handed couplings.
The analysis assumes a top-quark mass of $m_t$ = 172.5~GeV/$c^2$, consistent with the world average value~\cite{mtop}.
%


We use a data sample enriched in 
$\ttbar \to W^+ b W^- \bar{b} \to \ell \nu b q \bar{q}' \bar{b}$ events, 
where one of the $W$ bosons decays into quark pairs and the other into lepton pairs. The data was acquired by the Collider Detector at Fermilab (CDF II)~\cite{c:cdfDet}, which recorded $\ppbar$ collisions from Fermilab's Tevatron operating at $\sqrt{s}=1.96$~TeV. Most of the events used in the analysis were collected using inclusive-lepton online event selections (triggers) that required a high-transverse-momentum ($p_T$) electron or muon in the central (pseudorapidity $|\eta|<1.1$) detector region~\cite{met}.  The acceptance for $\ttbar$ events is increased by also using muon events satisfying a trigger that requires large missing transverse energy $\met$~\cite{met} with either an energetic electromagnetic cluster or two separated jets~\cite{trigger} ($\met$+jets trigger). After all data quality requirements, the sample collected corresponds to an integrated luminosity of 8.7~$\rm{fb}^{-1}$.


Candidate events are required 
to have a single isolated electron or muon candidate with $E_T>20$~GeV;
missing transverse energy $\met >20$~GeV consistent with expectations  from the undetectable high energy neutrino; and at least four energetic 
jets with $E_T>20$~GeV and $|\eta|<2$. Jets are reconstructed using a cone algorithm~\cite{MidPoint} with radius $\Delta R=0.4$ in $\eta-\phi$ space,
and their energies are corrected to the particle level by accounting for detector-response non-uniformities as a function of jet $\eta$, for effects from multiple $p \bar{p}$ interactions, and for the hadronic jet energy scale of the
calorimeter~\cite{jesnim}.  At least one jet must be identified
as having originated from a $b$ quark ($b$ tag) using an algorithm that 
exploits the long lifetime of $b$ hadrons and their large boost from the 
decay of the top quark.  We require decay vertices 
displaced from the primary $\ppbar$-interaction vertex~\cite{secvtx}.

The backgrounds to the $\ttbar$ signal are from multi-jet production 
(QCD), direct $W$-boson production in association with jets ($W$+jets), and 
electroweak backgrounds (EWK) composed of diboson ({\it WW}, {\it WZ}, 
{\it ZZ}) and 
single top-quark production.  The $W$+jets background includes events with 
$b$-tagged $b$-quark jets as well as erroneously $b$-tagged light-flavor or charm-quark jets. 
The method for estimating the background is described in detail
in Ref.~\cite{xsection}.   Table~\ref{tab:events} shows the
expected sample composition using a $\ttbar$ cross section of 
7.4~pb from Ref.~\cite{ttXSTheory}.
Events that satisfy one of the high-$p_T$ lepton triggers and the 
$\met$+jets trigger are assigned to the lepton-triggered sample and removed from the $\met$+jets-triggered sample.


\begin{table}[h]
\caption{Number of expected and observed events in $8.7$ fb$^{-1}$ of
data for each of the exclusively-defined samples. A $\ttbar$ cross section of 7.4~pb is assumed for the signal.}
\begin{ruledtabular}\begin{tabular}{lccc}
 Process  & Central     &  Central   &  $\met$+jets \\
          & $e$         &    $\mu$   &   $\mu$      \\ \hline
\vspace*{-0.3cm}\\
$\ttbar$  & 923 $\pm$ 93 & 696 $\pm$ 54   &   441 $\pm$ 44 \\
$W$+jets  & 160 $\pm$ 41 & 125 $\pm$ 19   &   106 $\pm$ 21 \\
EWK       &  36 $\pm$ 17 &  27 $\pm$ 11   &    16 $\pm$  8 \\
QCD       & 121 $\pm$ 48 &   6 $\pm$  2   &     7 $\pm$  3  \\
\hline
Total expected    &  1239 $\pm$ 113 &  853 $\pm$ 59  & 569 $\pm$ 50 \\
\hline
Observed  &  1226  & 804  &  544 \\
\end{tabular}
\end{ruledtabular}
\label{tab:events} \rm
\end{table}

\begin{figure}[htb]
\centering
\includegraphics[width=0.5\textwidth]{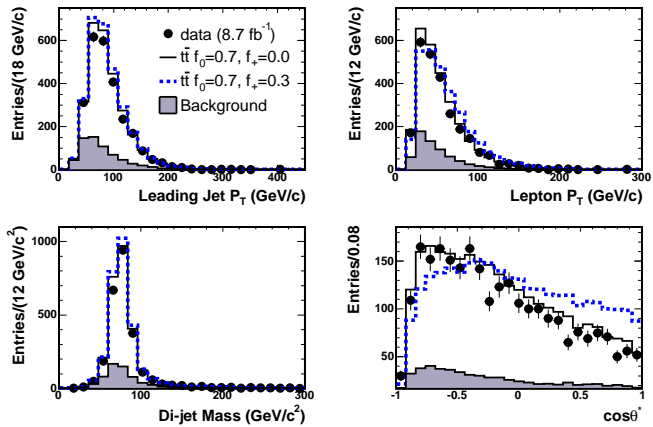}
\hfill
\caption{\label{fig:dataVal} The data (points) are compared to the MC
prediction for different $W$ polarization fractions.  The background 
predictions are shown as the shaded histograms while the 
signal-plus-background predictions are shown as the open histograms corresponding to ($f_0$, $f_+$) values of (0.7, 0.0) and (0.7, 0.3)
for the solid and dashed lines, respectively.  The four kinematic variables 
displayed are the leading jet $p_T$, the lepton $p_T$, the invariant mass
of the pair of light-quark jets from the hadronically decaying $W$ boson,
and the $\Cts$ of the leptonically decaying $W$ boson.  For the latter two
distributions the jet-parton assignment most consistent with the signal 
hypothesis is shown.
}
\end{figure}

The $\ttbar$ signal events are modeled using the {\sc herwig}~\cite{herwig} 
Monte Carlo (MC) generator.
The QCD background is modeled using data control samples.
The {\sc alpgen}~\cite{alpgen}, {\sc madevent}~\cite{Madevent} and 
{\sc mc@nlo}~\cite{MC@NLO} programs, with {\sc pythia}~\cite{pythia} or {\sc herwig} supplying the parton-shower and fragmentation model, and the
full {\sc pythia}~\cite{pythia} generator, are used to model the remaining backgrounds and for estimating systematic uncertainties.  
A {\sc geant}-based simulation~\cite{geant} is used to model the response of the CDF~II detector for these simulated samples.
The signal and background modeling has been extensively checked.
The observed data and the predicted signal-plus-background distributions 
for various kinematic variables are compared in Fig.~\ref{fig:dataVal}.  
We further validate the background model using a high-statistics 
background-dominated data control sample obtained vetoing events with a
$b$-tagged jet.

To determine the polarization fractions $f_0$, $f_-$, and $f_+$, 
an unbinned likelihood technique 
is employed.  The likelihood is calculated using the theoretical matrix 
elements for both the dominant signal process, $q\bar{q} \rightarrow \ttbar
$, and the main background process, inclusive production of $W$+jets.  
The method assumes that $\ppbar\to\ttbar$ production is accurately described by the SM and includes the physical constraint $\sum_{i}^{0,-,+} f_{i} = 1$.  
The technique was first developed for measuring the mass of the top quark and for determining $f_0$ when constraining $f_+$ to its SM value~\cite{ME}.  We have extend the technique to enable the simultaneous determination of 
$f_0$ and $f_+$~\cite{CDF2.7ifb}. 
The \ttbar\ matrix element is expressed in terms of the $W$-boson polarization fractions and the cosine of the angle $\theta^*$ between the momentum of the charged lepton or down-type quark from the $W$-boson decay in the $W$-boson rest frame and the direction of the top quark.
For the signal $q\bar q\rightarrow t\bar t$ process~\cite{mahlonparke},
the leading-order matrix element is used,
\begin{equation*}
  |M|^2 = \frac{g_s^4}{9} F_{\ell} \bar{F}_{h} (2 - \beta^2 \mathrm{sin}^2\theta_{qt}) \mathrm{,}
  \label{eqn:ME} \nonumber
\end{equation*}
where $g_s$ is the strong coupling constant, $\theta_{qt}$
describes the angle between the incoming parton and the top quark in
the rest frame of the incoming partons, and $\beta=v/c$ where $v$ is the velocity of the top quarks in the same rest frame.
The factors $F_{\ell}$ and $\bar{F}_{h}$ correspond to top quarks with 
a leptonic and a hadronic $W$-boson decay, respectively, such that
\begin{eqnarray*}
  F_{\ell}  &=& \frac{ 2 \pi g_W^4 m_{\bar{\ell}\nu}^2 }{3 m_t \Gamma_t}
  \left( 2 E_{b}^{*2} + 3 E^*_{b} m_{\bar{\ell}\nu} + m_{b}^2 \right) \\
  & & \left[ \frac{3}{8} \left( 1 + \Cts \right)^2 f_+
     + \frac{3}{4} \left( 1 - \CtsSq \right) f_0 \right. \\
  & & \left. \mbox{} + \frac{3}{8} \left(1- \Cts\right)^2 
        \left(1-f_0 - f_+ \right)\right] \mathrm{,}
  \label{eqn:lepFCts}
\end{eqnarray*}
where $g_W$ is the weak coupling constant,
$m_{\bar{\ell}\nu}$ is the charged lepton-neutrino invariant mass, 
$\Gamma_t$ is the decay width of the top quark, and
$ E^{*}_{b} =  \frac{ m_t^2 - m_{b}^2 - m^2_{\bar{\ell}\nu}}{2 m_{\bar{\ell}\nu}} $.
The hadronic factor ${F}_{h}$ is similar. Since we cannot distinguish between up-type and down-type quark jets from the hadronic $W$-boson decay, both jet-quark assignments are used and
an average $\bar{F}_{h}$ is calculated from the two permutations. 
The background matrix element is approximated using the sum of $W$+jets matrix elements from the {\sc vecbos}~\cite{vecbos} program.

The polarization fractions are determined by maximizing the likelihood function $L$ with respect to $f_0$, $f_+$, and the fraction of events consistent with the $\ttbar$ signal hypothesis, $C_s$,  
\begin{equation*}
\hspace*{-0.65mm}
  L(f_0, f_+, C_s|x) = \prod_{i=1}^{N} \left[C_{s} \frac{P_{s}(x_{i};f_0, f_+)}{\langle A_s(x_{i};f_0, f_+) \rangle} + (1 - C_s) \frac{P_{b}(x_{i})}{\langle A_b(x_{i}) \rangle }\right]\mathrm{,}
\label{eq01}
\end{equation*}
where $N$ is the number of observed events, $x$ is the vector of observed momenta of the final state partons, and $\langle A_s \rangle$ and $\langle A_b \rangle$ 
are the average acceptances for $\ttbar$ and $W$+jets background events, respectively.  The dependence of $\langle A_s \rangle$ on the polarization fractions is properly included.
The signal probability density, $P_s$, and background probability density, $P_b$, are calculated as described in Ref.~\cite{MEAT} and integrated over the relevant differential cross section, which depends on the matrix 
elements described above, convolved with the proton parton distribution functions (PDFs).  Poorly-known parton-level quantities are integrated 
out. The parton four-momenta are estimated from the measured momenta of the 
trigger lepton and the four highest-transverse-energy $E_T$ jets in the event.  Detector resolution effects are accounted for with transfer functions derived from simulated $\ttbar$ samples.
There is an ambiguity in the jet-parton assignments and all permutations are used for each event.  When calculating $P_s$ we fix the top-quark mass to $m_t = 172.5$~GeV/$c^2$ and scan the $\left(f_0,~f_+\right)$ parameter space. The calculation of $P_b$ is independent of $m_t$, $f_0$, and $f_+$.



The polarization fractions determined from the likelihood fit differ from the true polarization fractions because the signal and background probability densities contain approximations.  For example, they do not accurately account for the effects of extra jets arising from initial and final state radiation (ISR/FSR) or for the full set of contributing background processes.  Thus, the results of the likelihood fit are calibrated with samples of $\ttbar$ and background events simulated using the sample composition of Table~\ref{tab:events} and assuming a variety of input $\left(f_0,~f_+\right)$
values. The mean measured polarization fractions determined from the simulated samples are plotted against the true polarization fractions and a calibration function is determined from a linear fit to the resulting curve.  For the one-dimensional polarization measurements a one-dimensional
calibration function is employed, while for the simultaneous determination of $( f_0 , f_+ )$ a two-dimensional calibration function is used.  
The resulting calibration functions are used to estimate the
true polarization fractions from the measured polarization fractions 
determined from the three separate likelihood fits.
The uncertainties on the coefficients of the calibration functions
are included in the method-related systematic uncertainties.
Even though the likelihood can be calculated only for the physical values of $f_0$ and $f_+$,
after calibration the corrected measured values can be slightly outside their physical ranges. 


The robustness of the fitting and calibration procedure is tested over all physical values of ($f_0$,$f_+$) using simulated experiments 
constructed to have the number of observed data events and the sample composition of Table~\ref{tab:events}.  No significant biases are 
observed.  However, near the physical boundaries the statistical uncertainty is underestimated by as much as a factor of two.  A correction 
to the statistical uncertainty is applied in these regions.
Assuming the SM, the expected statistical uncertainties
for the simultaneous measurement, after all corrections,  
are $\pm0.075$ and $\pm0.047$ for $f_0$ and $f_+$, respectively.


The sources of systematic uncertainty affecting the measurements
are summarized in Table~\ref{tab:SystSummary}.
All systematic uncertainties are determined by performing simulated experiments in which the systematic parameter in question is varied,
the default method and calibrations are applied, and the shifts in the mean
measured polarization fractions are used to quantify the uncertainty.
All shifts are evaluated using the SM polarization fractions.
The dominant source of systematic uncertainty for the simultaneous 
measurement of $f_0$ and $f_+$ is due to uncertainties on the background shape and normalization.  For the model-dependent measurements several 
sources of systematic uncertainty contribute at a comparable level.

\noindent 
\begin{table}[h]
\caption[Summary of systematic uncertainties.]{Summary of systematic uncertainties.}
\begin{ruledtabular}\begin{tabular}{lcccc}
Source            & $\Delta f_0$  & $\Delta f_+$
                  & $\Delta f_0$  & $\Delta f_+$ \\ 
                  &               &     & \multicolumn{2}{c}{Simultaneous} \\ \hline 
Background model   & 0.007  & 0.011  & 0.049  & 0.036  \\ 
ISR/FSR            & 0.011  & 0.017  & 0.022  & 0.023  \\ 
MC generator       & 0.012  & 0.009  & 0.023  & 0.011  \\ 
Color reconnection & 0.013  & 0.010  & 0.020  & 0.016  \\ 
Method-related     & 0.014  & 0.020  & 0.018  & 0.016  \\ 
Jet energies       & 0.016  & 0.017  & 0.010  & 0.022  \\ 
PDF                & 0.024  & 0.013  & 0.009  & 0.016  \\ 
Multiple interactions  
                   & 0.009  & 0.013  & 0.008  & 0.014  \\ 
\hline
Total              & 0.040  & 0.040  & 0.067  & 0.058  \\ 
\end{tabular}
\end{ruledtabular}
\label{tab:SystSummary}
\end{table}

The uncertainty on the background model is determined by simulating experiments with the mean number of total background events increased and decreased by one standard deviation while keeping the relative contributions of the various background sources fixed as given in 
Table~\ref{tab:events}.  Half the mean difference between these two sets of simulated experiments is assigned as the systematic uncertainty. The background model uncertainty also adds in quadrature the largest observed change when varying the normalization for each background source in turn by one standard deviation, while keeping the total background fixed thereby 
affecting the shape of the background distributions.
The ISR/FSR uncertainty is evaluated using MC samples generated with 
ISR/FSR settings that are amplified or damped relative to the default settings.
The MC generator uncertainty is evaluated by comparing between $\ttbar$ MC generated by {\sc pythia} and {\sc mc@nlo} with parton showering done by {\sc herwig}; it includes uncertainties from not using the NLO matrix element in the generator, choice of parton shower model and modeling of 
$\ttbar$ spin-correlation. In Ref.~\cite{CDF2.7ifb} we only listed uncertainty from choice of parton shower model.
The color reconnection systematic uncertainty~\cite{ColorReconnection} is evaluated using MC samples generated with and without color reconnection effects adopting different configurations~\cite{PythiaTunes} of {\sc pythia}.
The method-related uncertainty includes propagating the uncertainty 
on the fit parameters of the calibration functions, including their correlations.  It also includes the uncertainties related to the statistics of the MC samples used to perform the calibration.
The uncertainties in the jet energy scale corrections are propagated through the analysis by varying the corrections within one standard deviation and recording the resulting shifts in the polarization fractions.
Variations associated with the choice of PDF and their uncertainties affect the $\ttbar$ acceptance and are included as a systematic uncertainty.
The luminosity profile of the MC samples does not exactly match that of the data.  The associated systematic uncertainty is evaluated by varying the MC distribution of events containing multiple $\ppbar$ interactions so that it matches the data distribution.


Using the 2574 data events that meet all selection criteria we perform three measurements of the $W$ boson polarization fractions.
For the model-independent measurement we simultaneously determine $f_0$ and $f_+$ to be
\begin{eqnarray*}
   f_{0} & = & \:0.726 \pm 0.066 \: (\rm{stat}) \pm 0.067\: (\rm{syst}) \\
   f_{+} & = &  -0.045 \pm 0.044 \: (\rm{stat}) \pm 0.058\: (\rm{syst})
\end{eqnarray*}
after all corrections. The correlation between $f_0$ and $f_+$ is $-0.69$. 

\begin{figure}[htb]
\centering
\includegraphics[width=0.35\textwidth]{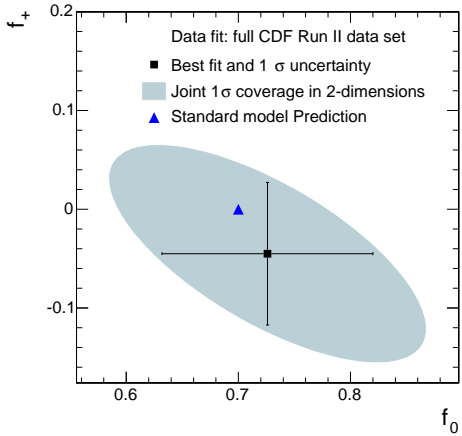}
\hfill
\caption{\label{fig:fig-test-CL-unCorrGrid-morePSE}
The final results for the simultaneous measurement of $f_0$ and $f_+$ 
after all corrections and including statistical and systematic uncertainties.  
The square marker with error bars shows the point of maximum likelihood 
and the one-dimensional $1\sigma$ uncertainties obtained from the 
$-\ln(L_{\rm{syst}}/L_{\rm{syst}}^{\rm{max}}) = 0.5$ contour. The shaded ellipse corresponds to the two-dimensional $1\sigma$ uncertainty obtained from the 
$-\ln(L_{\rm{syst}}/L_{\rm{syst}}^{\rm{max}}) = 1.15$ contour. The 
triangle marker shows the SM prediction.
}
\end{figure}

The two-dimensional likelihood contour obtained from the data only includes the statistical uncertainty. The final contour, including statistical and 
systematic uncertainties, is obtained by analytic convolution of the 
data likelihood with a two-dimensional Gaussian representing the systematic uncertainties, resulting in a new likelihood $L_{\rm{syst}}$. 
Figure~\ref{fig:fig-test-CL-unCorrGrid-morePSE} shows the point estimate with error bars, corresponding to one-dimensional 68.27\% confidence level (C.L.) uncertainties, and the two-dimensional 68.27\% C.L. region, obtained using $-\ln(L_{\rm{syst}}/L_{\rm{syst}}^{\rm{max}}) = 0.5$ and $1.15$, respectively, where 
$L_{\rm{syst}}^{\rm{max}}$ is the maximum value for the likelihood 
$L_{\rm{syst}}$. We estimate a shift of $\mp(0.010 \pm 0.004)$ in $f_0$ and 
$\pm(0.012 \pm 0.002)$ in $f_+$ per $\pm1$ GeV/$c^{2}$ shift in the 
top-quark mass from the central value of 172.5 GeV/$c^{2}$.

For the measurement fixing $f_+=0$, we obtain after all corrections
$f_{0} =  0.683 \pm 0.042\: (\rm{stat}) \pm 0.040\: (\rm{syst})$.
For the measurement fixing $f_0=0.70$, we measure after all corrections 
$f_{+} = -0.025 \pm 0.024\: (\rm{stat}) \pm 0.040\: (\rm{syst})$.
We estimate a shift of $\pm(0.007 \pm 0.002)$ 
in $f_0$ and $\pm(0.008 \pm 0.001)$ in
$f_+$ per $\pm1$ GeV/$c^2$ shift in the top-quark mass from the central value of 172.5 GeV/$c^2$.

In summary, we present measurements of the polarization of $W$ bosons in 
top-quark decays using the lepton-plus-jets final state and the full CDF Run~II data sample corresponding to an integrated luminosity of
8.7~$\rm{fb}^{-1}$.  A matrix-element technique is used to significantly improve the statistical precision relative to
previously-used techniques and is extended to allow for a 
simultaneous determination of $f_0$ and $f_+$ in a model-independent manner.  This result improves the statistical precision 
on both the model-independent and model-dependent determinations 
of $f_0$ and $f_+$ by a factor of 1.6 compared to the previous CDF measurement~\cite{CDF2.7ifb} in the lepton+jets channel.
Our result is the first $W$-polarization measurement using the full data set from Tevatron Run II, and is the most precise single-channel measurement to date from the Tevatron.  The results from the model-independent and model-dependent measurements of  $W$ polarization are limited by the size of the systematic uncertainties,
and have a precision comparable to the combination reported in Ref.~\cite{TevWhelComb}. All the results are consistent with the SM.

We thank the Fermilab staff and the technical staffs of the participating institutions for their vital contributions. This work was supported by the U.S. Department of Energy and National Science Foundation; the Italian Istituto Nazionale di Fisica Nucleare; the Ministry of Education, Culture, Sports, Science and Technology of Japan; the Natural Sciences and Engineering Research Council of Canada; the National Science Council of the Republic of China; the Swiss National Science Foundation; the A.P. Sloan Foundation; the Bundesministerium f\"ur Bildung und Forschung, Germany; the Korean World Class University Program, the National Research Foundation of Korea; the Science and Technology Facilities Council and the Royal Society, UK; the Russian Foundation for Basic Research; the Ministerio de Ciencia e Innovaci\'{o}n, and Programa Consolider-Ingenio 2010, Spain; the Slovak R\&D Agency; the Academy of Finland; and the Australian Research Council (ARC).

\bibliographystyle{unsrt}

\begin{thebibliography}{99}
\bibitem{ChargeConjugate} Charge-conjugate modes are included implicitly throughout this paper.
\bibitem{met}
We use a cylindrical coordinate system where the $z$ axis is
along the proton beam direction, $\phi$ is the azimuthal angle, and 
$\theta$\ is the polar angle.
Pseudorapidity is $\eta = -\ln\tan(\theta/2)$, while transverse momentum is 
$p_T = |p| \sin \theta$, and transverse energy is $E_T = E \sin \theta$.
Missing transverse energy, \met, is defined as the magnitude of 
$-\sum_iE_T^i \hat{n}_i$, where $\hat{n}_i$\ is the unit vector in the 
azimuthal plane that points from the primary \ppbar\ collision vertex to
the $i$th energy deposition in the CDF~II calorimeter.
\end{thebibliography}

\end{document}